\documentclass{iaus}
\usepackage{graphicx}

\title[New Solar System] 
{Icy Bodies in the New Solar System}

\author[David Jewitt]   
{David Jewitt$^1$}

\affiliation{$^1$Dept. Earth and Space Sciences and Institute for Geophysics and
Planetary Physics, UCLA, \\
3713 Geology Building, 595 Charles Young Drive East, Los Angeles, CA 90095-1567}

\pubyear{2010}
\volume{xxx}  
\pagerange{1--14}
\jname{IAU General Assembly 2009}
\editors{Julio Fernandez et al., Editors.}
\begin{document}

\maketitle

\begin{abstract}
This brief paper summarizes a ``key general review'' with the same title given at the IAU meeting in Rio de Janeiro. 
The intent of the review talk was to give a broad and well-illustrated overview of recent work on the icy middle and outer Solar system, in a style interesting for those astronomers
whose gaze is otherwise drawn to more distant realms.  The intent of this written review is the same.

\end{abstract}

\firstsection 
\section{Introduction}
The last 20 years have seen an incredible burst of research on the small bodies of the Solar system, particularly addressing the icy objects in its middle (from Jupiter to Neptune) and outer (beyond Neptune) parts.  This burst has been driven largely by ground-based telescopic surveys, revealing previously unknown populations in regions formerly thought to be empty.  Through physical observations with the world's largest telescopes, the characters of many known icy bodies have also been more firmly established.  Separately, dynamicists have greatly added to our appreciation of the complexity of the Solar system through their clever exploitation of ever-faster and cheaper computers and the need to fit new observations.  Lastly, several small bodies (comets, asteroids and planetary satellites) have been approached by spacecraft, providing invaluable close-up data on objects previously quite beyond human reach.   

As a result of all this activity, many of us have come to realize that the small bodies, although they contain a negligible fraction of the total mass in the Solar system, in fact carry a disproportionately large fraction of the scientifically useful information.  They are macroscopic analogs of the radioactive elements (which, although they are mass-wise insignificant play a central scientific role as our best cosmic chronometers).  This is especially true concerning information about the origin and evolution of the Solar system, for two reasons.  Firstly, many of the small bodies have escaped substantial thermal alteration since the formation epoch.  Their chemical and molecular constitutions therefore approach the initial conditions, or at least approach them much more closely than do large bodies like the Earth (in which the gravitational binding energy is so large as to have caused melting at accretion).  Secondly, the small bodies are so numerous, and their orbits so accurate, that they can be used to map dynamical parameter space and to trace processes occurring in the protoplanetary disk and after.  

\begin{figure}[]
\begin{center}
 \includegraphics[width=4.4in]{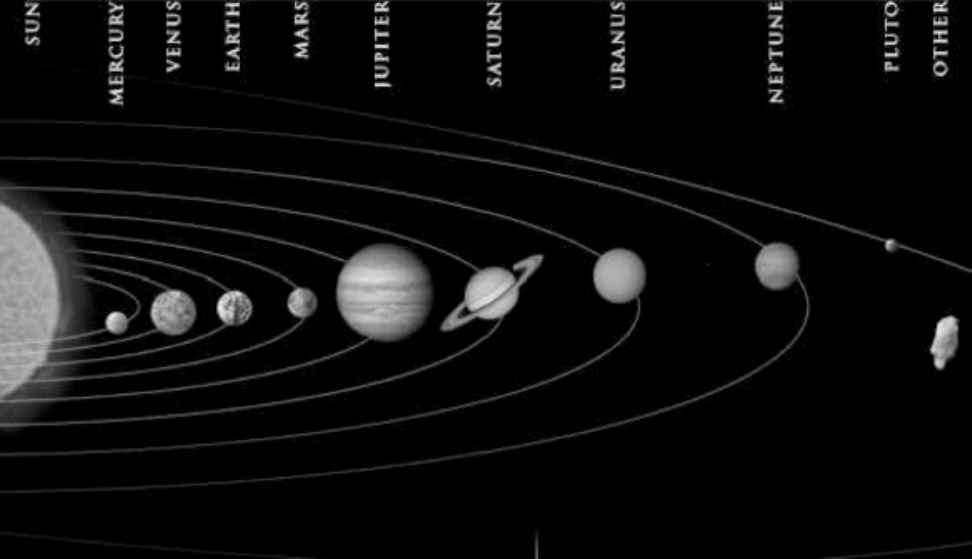} 
\end{center} 
\caption{Your grandmother's Solar system: outside the Sun, 9 objects worthy of names and a grab-bag full of insignificant things called ``Other''.  Recent research into ``Other'' has yielded incredible scientific treasure out of all proportion to the tiny mass of the objects therein.  Figure courtesy NASA.}
\label{fig1}
\end{figure}

Figure \ref{fig1} emphasizes the resulting change in perception of the Solar system in a visual way.  It illustrates the old view of the Solar system, in which there are nine important objects outside the Sun, and a host of smaller, relatively unimportant bodies that have been unceremoniously thrown into a bin called ``Other''.   The irony of the new Solar system is that it is precisely these throw-away  ``Other'' objects which have provided much of the scientific excitement and sense of renewal in our field.  They are the objects of this review.


In this review, I will focus on the main reservoirs of icy objects in orbit about the Sun.  These are, in the order of their discovery,
the Oort cloud, the Kuiper belt and the outer asteroid belt.  First, though, it is worth having a look at some of the objects of our attention.

\begin{figure}[b]
\begin{center}
 \includegraphics[width=4.4in]{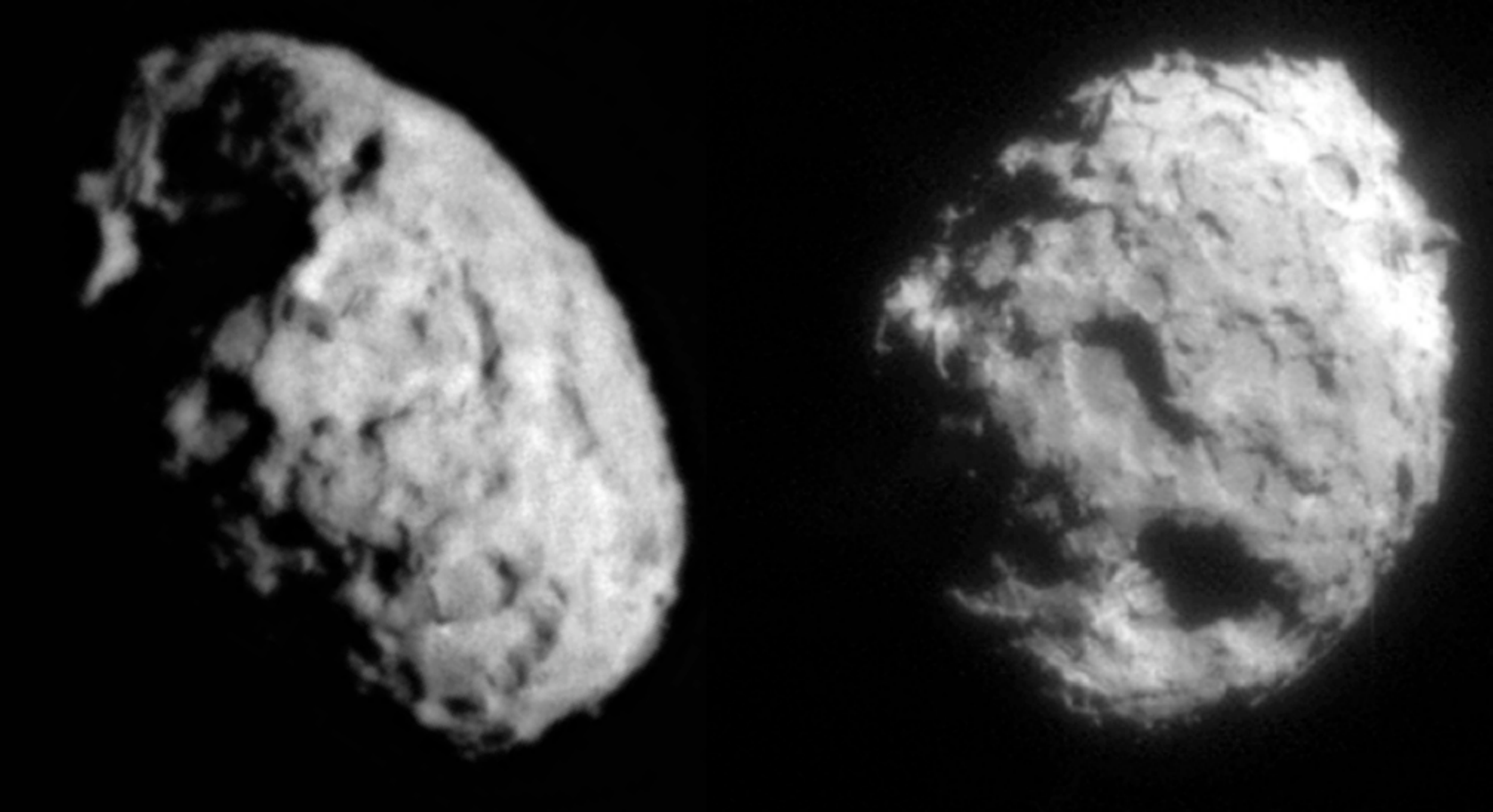} 
\end{center} 
\caption{Nucleus of comet P/Wild 2, a recently escaped Kuiper belt object.  The density of this $\sim$4 km diameter nucleus is not known, but believed to be less than
1000 kg m$^{-3}$, consistent with an ice-rich, porous make-up.  Craters in the surface are
probably not impact craters, but may result from mass-loss driven by sublimation.  Image courtesy NASA's Stardust mission.}
\label{wild2}
\end{figure}

\subsection{Photo-interlude}
Figure \ref{wild2} shows in-situ images of the nucleus of Jupiter family comet comet 81P/Wild 2, taken from two different perspectives by NASA's Stardust mission.  This is a $\sim$4 km scale icy object whose orbit is consistent with a source in the Kuiper belt.  Although it is icy, the albedo of the nucleus is very small ($\sim$0.06), probably because the ices in this bodies are buried in a refractory, perhaps carbon-rich mantle.   The numerous depressions in the surface of P/Wild 2 at first glance resemble impact craters, but closer examination suggests that they are more likely to be related to the outgassing phenomenon.  They are, frankly, not well understood.

\begin{figure}
\begin{center}
 \includegraphics[width=3.4in]{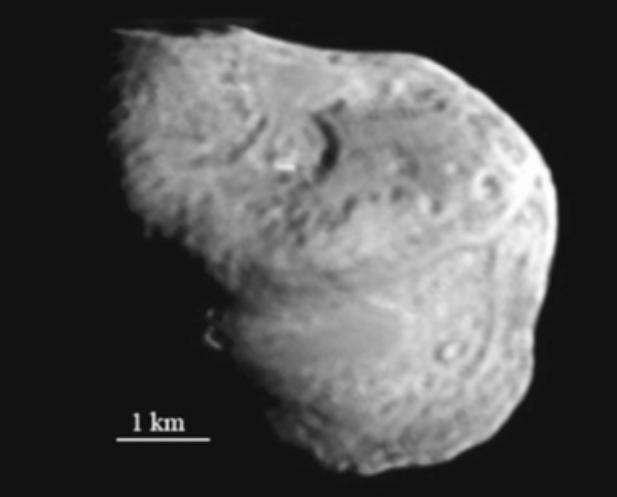} 
\end{center} 
\caption{Nucleus of comet P/Tempel 1.  The nucleus is irregular, with $\sim$6 km effective diameter, while the bulk density is uncertain, but lies in the range 200 to 1000 kg m$^{-3}$. Note the large tongue-shaped smooth region in the lower half of the nucleus (a ground-hugging flow on an object whose escape velocity is $\sim$1 m s$^{-1}$?) and the relative lack of large craters compared to Figure \ref{wild2}.  Image courtesy NASA's Deep Impact mission.}
\label{tempel1}
\end{figure}

A different Jupiter family comet (Kuiper belt source) nucleus is shown in Figure \ref{tempel1}.  While similar in size to P/Wild 2, this 
nucleus is less densely cratered and a large swath of the surface is occupied by a smooth, lobate flow-like feature.  Again, the surface geology is a mystery for which several solutions have been proposed.  The bigger mystery is why these two JFC nuclei should look so different.

\begin{figure}
\begin{center}
 \includegraphics[width=3.4in]{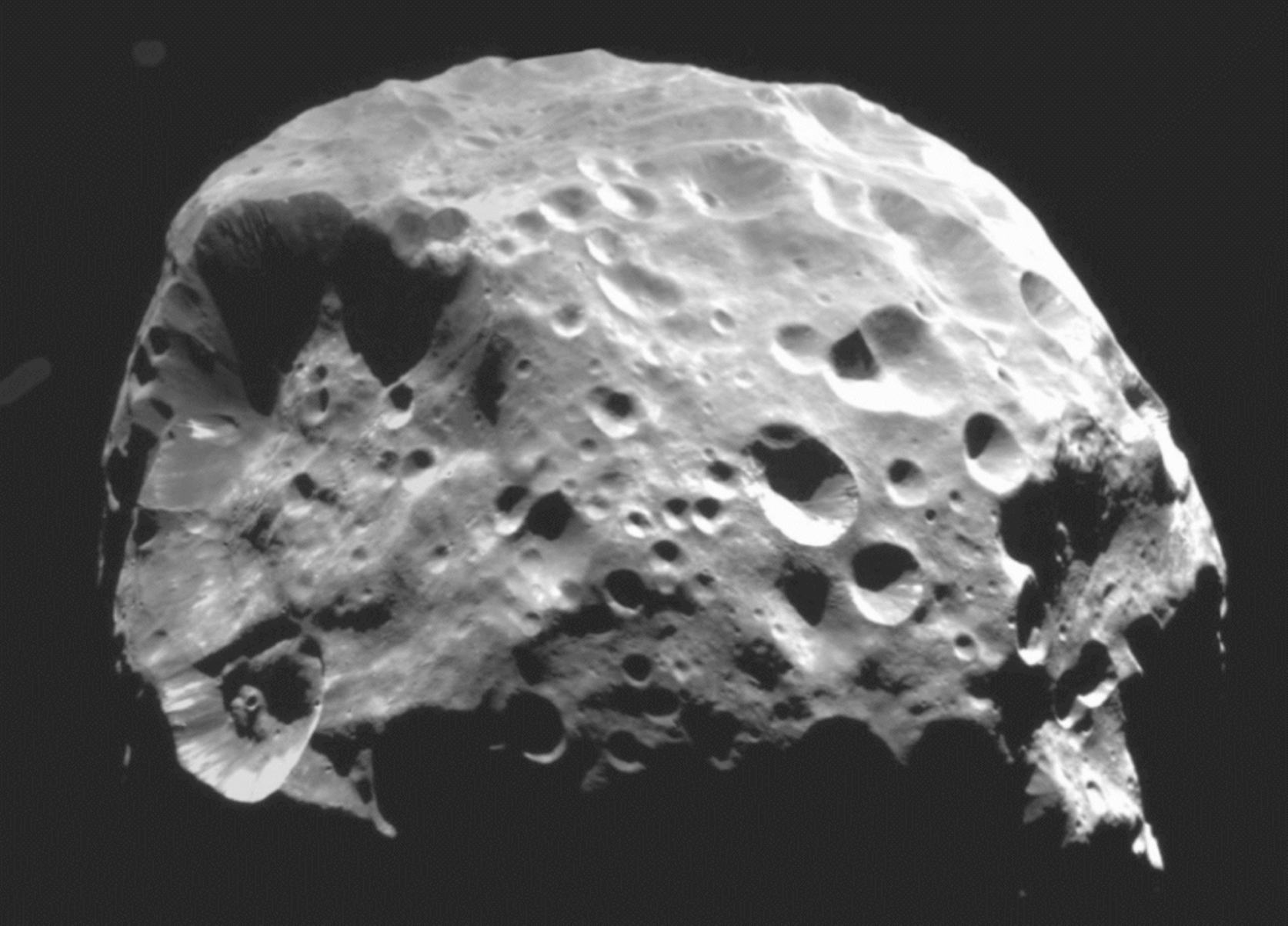} 
\end{center} 
\caption{Saturnian satellite Phoebe, $\sim$220 km in diameter.  Image courtesy NASA's Cassini mission.}
\label{phoebe}
\end{figure}

Figure \ref{phoebe} shows Saturn's large irregular satellite, Phoebe.  With diameter $\sim$220 km, it is nearly two orders of magnitude
larger than the cometary nuclei shown in the previous figures.  The density is about 1630$\pm$45 kg m$^{-3}$ (Porco et al. 2005), requiring a rock-ice composition but admitting the possibility of considerable internal porosity.  The heavily cratered surface much more closely resembles the highly
cratered Lunar highlands, suggesting that the influence of near-surface ice (and its sublimation) is much less important on this body (as expected for its more distant location; Saturn is at 9.5 AU from the Sun and far too cold for water ice to sublimate).  Ices are identified in spectra of the surface.  The significance of Phoebe is that it, and other irregular satellites, might have been captured by the planets following formation in the Kuiper belt.

\begin{figure}
\begin{center}
 \includegraphics[width=3.4in]{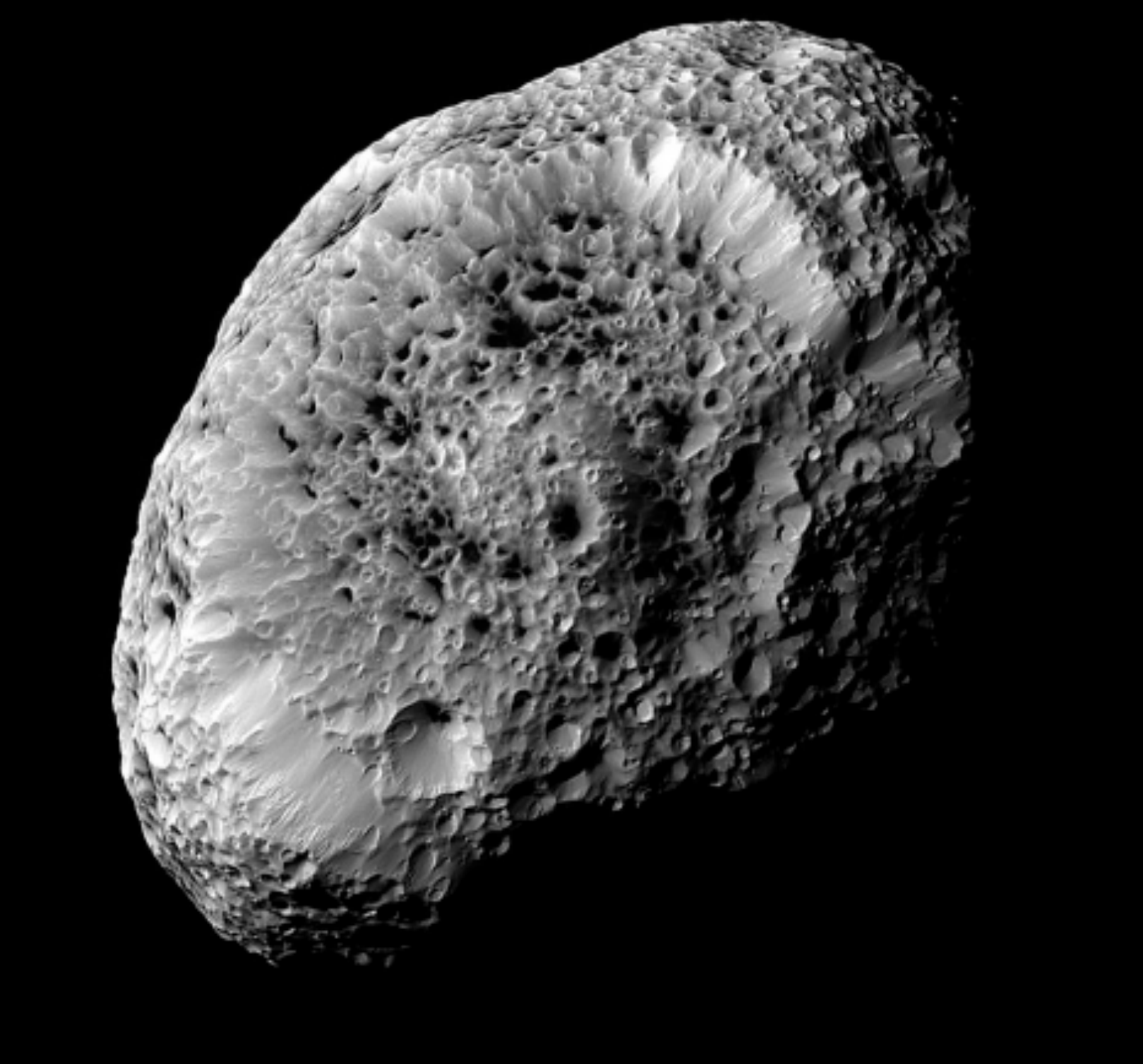} 
\end{center} 
\caption{The 400 km scale Saturnian satellite Hyperion, an extraordinarily porous body with a density only half that of water.  Image courtesy NASA's Cassini mission.}
\label{hyperion}
\end{figure}

The final Figure \ref{hyperion} shows Saturn's regular satellite Hyperion.  This large ($\sim$400 km scale) object 
is notable for several reasons.  It has an aspherical shape that allows planetary torques to imbue this satellite with a chaotic rotation.
It has a densely cratered ice-rich surface that is unlike any other so-far imaged.  Most of the camera-facing side of Hyperion in
Figure \ref{hyperion} is occupied by an impact crater with $\sim$20 km deep walls: the kinetic energy of the projectile that formed 
this crater must surely have been comparable to the energy needed to disrupt the satellite.  And the density of Hyperion is a remarkably low 
540$\pm$50 kg m$^{-3}$, indicating substantial porosity.  As a regular satellite, Hyperion probably formed in Saturn's accretion disk and has no likely connection with the major Sun-orbiting ice reservoirs.  Nevertheless, in its low density and highly battered surface it is probably representative of many objects in the outer Solar system that have yet to be imaged close-up by spacecraft.

\section{The Three Ice Reservoirs}
The three heliocentric ice reservoirs, the Oort cloud, the Kuiper belt and the main-belt, are the main topics of this review.  Other repositories of ice are also worthy subjects of intensive scientific study, including the
nuclei of comets, dead (or defunct) comets, the Damocloids (probably dead Halley-type comets), the planetary satellites and
the Trojans, both those of Jupiter and of Neptune.  Limited space forces me to suppress them in this article. 

The three reservoirs supply comets in three dynamical classes, the 
long-period comets (Oort cloud), the short-period (specifically ``Jupiter family'') comets (Kuiper belt) and the main-belt comets (outer asteroid belt). A simple way to think about the different comet types is in terms of their Tisserand parameters measured with respect to Jupiter (Vaghi 1973).  The parameter is

\begin{equation}
T_J = \frac{a_j}{a} + 2\left[ (1-e^2)\frac{a}{a_j}\right]^{1/2} \cos(i)
\end{equation}

\noindent in which $a_j$ and $a$ are the semimajor axes of Jupiter and of the object and $e$ and $i$ are the
eccentricity and inclination of the object's orbit (Jupiter is assumed to have $e_j$ = $i_j$ = 0 in this problem).
This parameter, $T_J$, is a constant of the motion in the restricted three-body problem.  In practice, it is related to
 the close-approach velocity relative to Jupiter.  
Jupiter has $T_J$ = 3, comets from the Oort cloud have $T_J <$ 2, comets from the Kuiper belt have 2 $\le T_J \le$ 3 and
main-belt comets have $T_J >$ 3.  The parameter is an ambiguous discriminant when applied too closely (e.g. is an
object with $T_J$ = 2.99 really different from one with $T_J$ = 3.00?) because the restricted
three-body problem is only an approximation, there are important perturbations from many planets, and so on.  But a
majority of comets can be meaningfully linked to their source regions through this single number, $T_J$.

\section{Oort Cloud}
\begin{figure}[t]
\begin{center}
 \includegraphics[width=3.4in]{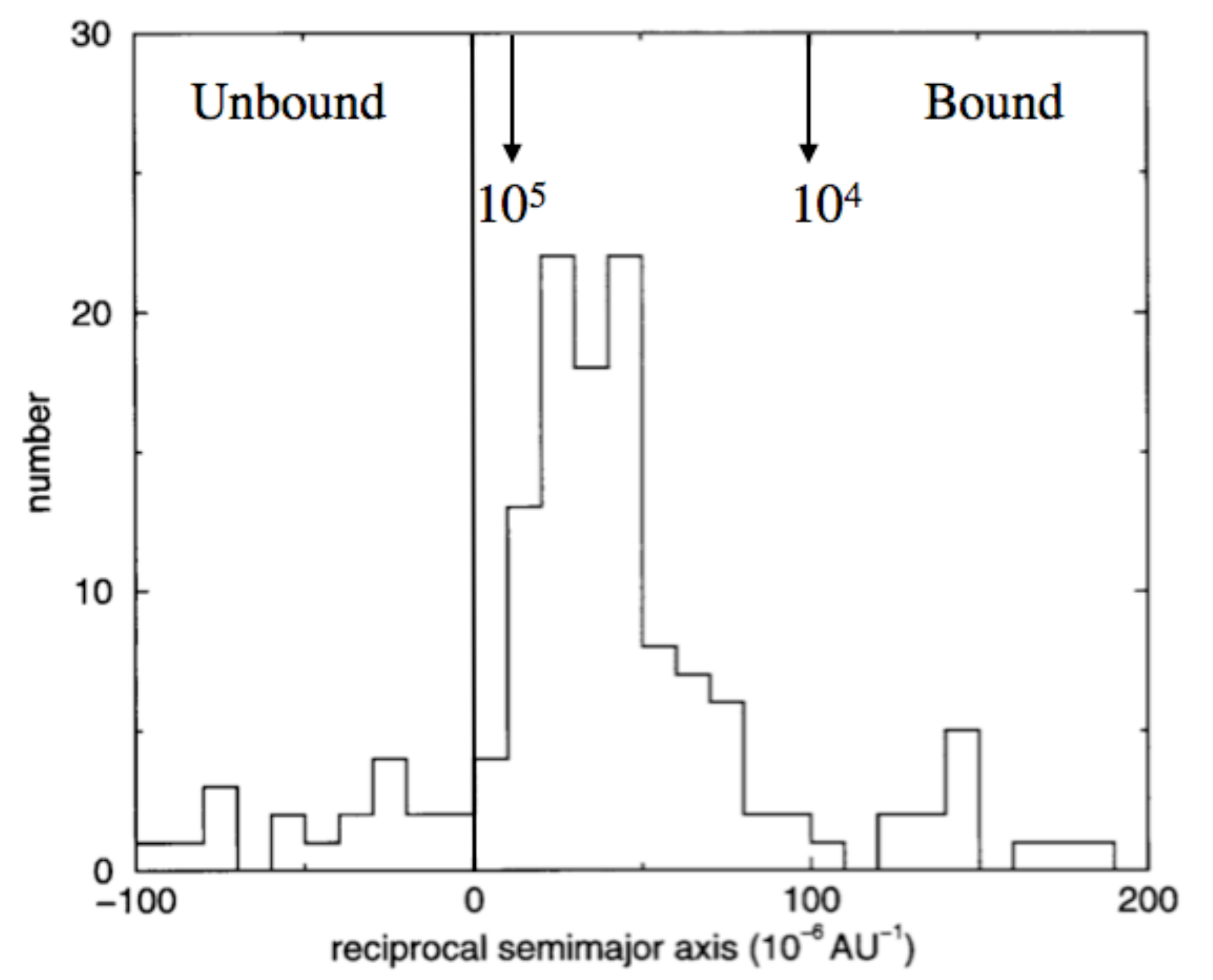} 
\end{center} 
\caption{Distribution of the reciprocal semimajor axes of comets.  Adapted from Marsden and Williams 1999. }
\label{oort}
\end{figure}

The principal evidence for the existence of the Oort Cloud lies in the distribution of the binding energies of long-period comets (Oort 1950).  The binding energies are proportional to the reciprocal semimajor axes, as plotted in Figure \ref{oort}.  The distribution is peaked to small, positive (i.e. gravitationally bound) values of $1/a$, corresponding to the large semimajor axes of comets falling into the Solar system from a distant, bound source.  On average, the energy of the orbit of an infalling comet will be slightly modified by interactions with the planets (note that the barycenter of the system lies near the photosphere of the Sun, rather than at its center, mostly as a result of the mass and 5 AU semimajor axis of Jupiter).   The key observation made by Oort is that the width of the peak in Figure \ref{oort} is small compared to the expected change in orbital energy from planetary scattering, meaning that comets in the peak have not previously passed through the Solar system many times.  The comets in the peak must include a large fraction of first-arrivals.  Over time, scattering in successive revolutions deflects comets out of the peak, either to be lost back to interstellar space or captured into more tightly bound orbits (larger $1/a$ in Figure \ref{oort}.  The shape of the Oort cloud is inferred to be spherical based on the (within observational limits) isotropic distribution of the arrival directions of long-period comets.  

Comets with $1/a <$ 0 in Figure \ref{oort} include objects that have been scattered into unbound trajectories by planetary perturbations during previous orbits, and those whose orbits have been
unbound by non-gravitational forces from anisotropic mass loss.  There are no known examples of the strongly hyperbolic comets that would be expected if comets entered the Solar system from interstellar space (Moro-Martin et al. 2009).

The estimated size and mass of the Oort cloud have been revised over the past half Century.  The cloud has no sharp edge, but the scale is 50,000 AU to 100,000 AU based on the best available comet orbit determinations.  The number of comets in the cloud is estimated by comparing the rate of arrival of new long period comets with the rate of erosion expected from external perturbations due to the combined action of passing stars and the torque from our own galaxy (Higuchi et al. 2007).  The estimate lies in the range 10$^{11}$ to 10$^{12}$ (Oort 1950, Francis 2002).  The mass in the Oort cloud is less certain still, since the size distribution of the long-period comet nuclei is very poorly known.  If all comets in the Oort cloud have radii of 1 km and densities equal to that of water, their combined mass would be in the range 0.1 M$_{\oplus}$ to 1 M$_{\oplus}$.   

The emplacement of the comets into the Oort cloud requires a two (or more) step process.  First, strong scattering by a growing planet leads to the ejection of comets from the protoplanetary disk of the Sun.  Those ejected at substantially less than the local escape speed fall back into the disk to be scattered again (or to collide with a growing planet).  Those ejected substantially faster than the local escape speed are lost forever into the interstellar medium.  Comets in a narrow range of ejection speeds are susceptible to the action of external perturbations from passing stars and from the galactic tide.  Their perihelia can be lifted out of the planetary domain and their orbits can be circularized to occupy the Oort cloud.  Estimates of the efficiency of emplacement into the Oort cloud vary.  Massive Jupiter (escape speed $\sim$60 km s$^{-1}$ compared with local escape speed from the Solar system $\sim$20 km s$^{-1}$) tends to launch comets into the interstellar medium.  Less massive Uranus and Neptune (escape speed $\sim$20 km s$^{-1}$, local Solar system escape speed $\sim$8 km s$^{-1}$) emplace comets into the Oort cloud more efficiently.  Overall, the eficiency for emplacement is in the range 1\% to 10\%.  If the mass of the Oort cloud is 0.1 to 1 M$_{\oplus}$, this means that 1M$_{\oplus}$ to 100M$_{\oplus}$ of comets must have been ejected during the process of planet formation.  

\subsection{Current Issues}
 \begin{itemize}

\item To fit the distribution of reciprocal semimajor axes (Figure \ref{oort}) Oort was forced to invoke a ``fading parameter'', whose purpose is to decrease the number of returning comets relative to the number observed.  In other words, dynamics alone would produce a flatter histogram for $1/a >$ 0 than is observed.  Oort suggested that freshly arriving comets might burn off a layer of ``interstellar frosting'' and so become fainter than expected on subsequent returns, thereby decreasing their number in any magnitude-limited plot.  The ``fading parameter'' is really a ``fudge parameter'' needed to make the model fit the data, and the physical nature of the fading remains unspecified, although suggestions abound (Levison et al. 2002, Dones et al. 2004).  It is tempting to speculate that the need for the fading parameter might be removed by some change in the dynamical part of the Oort cloud model but a recent careful study has failed to identify any such change (Wiegert and Tremaine, 1999, Dones et al. 2004).  Still, if the incoming long-period comets really fade, the important question remains ``why?''.

\item In Oort's model, the long-period comets arrive primarily from distant locations because the larger orbits are more susceptible to modification by tides from the Galaxy and from passing stars.  It has been suggested that the Oort cloud might contain a more tightly bound component at smaller mean distances (e.g. 5,000 AU instead of 50,000 AU), known as the Inner Oort Cloud (IOC).  These objects, if they exist, could provide a source from which to replenish the classical or Outer Oort Cloud (OOC).  It has been suggested that Kuiper belt object Quaoar is an IOC body but, with a perihelion of only 76 AU and a modest inclination and eccentricity, it seems more likely to have a dynamical connection to the Kuiper belt.  It has also been suggested that the IOC is the \textit{dominant} source of the long period comets and that, by implication, Oort's inferred much larger comet cloud is unimportant (Kaib and Quinn 2009).

Two big issues regarding the IOC remain unresolved.  First, and most importantly, does the IOC exist?  Precisely because this region is difficult to study observationally, we possess few serious constraints on its contents.  We have to admit that, observationally, large numbers of objects with considerable combined mass could exist in the IOC without violating any observational constraints.  But this does not mean that the objects are there.   Second, how was this region populated (if it was)?  The perturbations from passing stars and from the Galactic tide are too small to be responsible.  Instead, a densely populated IOC might be explained if the Sun formed in a dense star cluster, where the mean distance between stars was much smaller than at present and the magnitude and rate of perturbations were both much larger.  Formation in a cluster seems not unlikely: only a minority of stars form in isolation.  

\item Can we empirically constrain the mass that was ejected from the forming planetary system into the interstellar medium?  The orbit of a planet launching material from the Solar system should be greatly affected if the ejected mass rivals that of the planet, which sets a loose upper bound.  However, massive Jupiter (310 M$_{\oplus}$) anchors the system, and potentially very large masses could have been ejected.  
\end{itemize}

\section{Kuiper Belt}
The Kuiper belt is the region beyond Neptune ($a_j$ = 30 AU) in which vast numbers of small bodies have recently been discovered.  At first thought to be a simple, unprocessed remnant of the Sun's accretion disk, the Kuiper belt has been shaped in numerous ways by previously unsuspected evolutionary processes.  Significantly, the current mass is $\sim$0.1~M$_{\oplus}$, Jewitt et al. 1996, Trujillo et al. 2001), which is far too small for the known Kuiper belt objects to have grown by binary accretion on any reasonable timescale (Kenyon and Luu 1999).  The most likely explanation is that the modern-day Belt is the remnant of a (100$\times$ to 1000$\times$) more massive initial ring.   This primordial Kuiper belt could have shed its mass through some combination of dynamical processes (Gomes 2009) and collisional grinding.  

The velocity dispersion in the belt is $\Delta V \sim$ 1 to 2 km s$^{-1}$ (Jewitt et al. 1996, Trujillo et al. 2001), comparable to the gravitational escape velocity of the largest objects. With this $\Delta V$, collisions in the belt are primarily erosive, not agglomerative, leading to the idea that the Kuiper belt is eroding away.   It is reasonable to think of the Kuiper belt as the Sun's own ``debris disk'' (Wyatt 2008); a ring of parent bodies which collide to produce dust.  In fact, Kuiper belt dust has already been detected by the Voyager spacecraft (Gurnett et al. 1997, although the investigators apparently did not realize the source), leading to an estimate of the normal optical depth in dust $\tau_{kb}$ = 10$^{-7}$ (Jewitt and Luu 2000).  For comparison, the corresponding optical depths of the best-known (brightest) nearby debris disks are $\tau \sim$ 10$^{-3}$ to 10$^{-4}$ (Wyatt 2008).  If displaced to a nearby star, our own Kuiper belt would fall far below the thresholds of observational detection.  

\begin{figure}[t]
\begin{center}
 \includegraphics[width=3.8in]{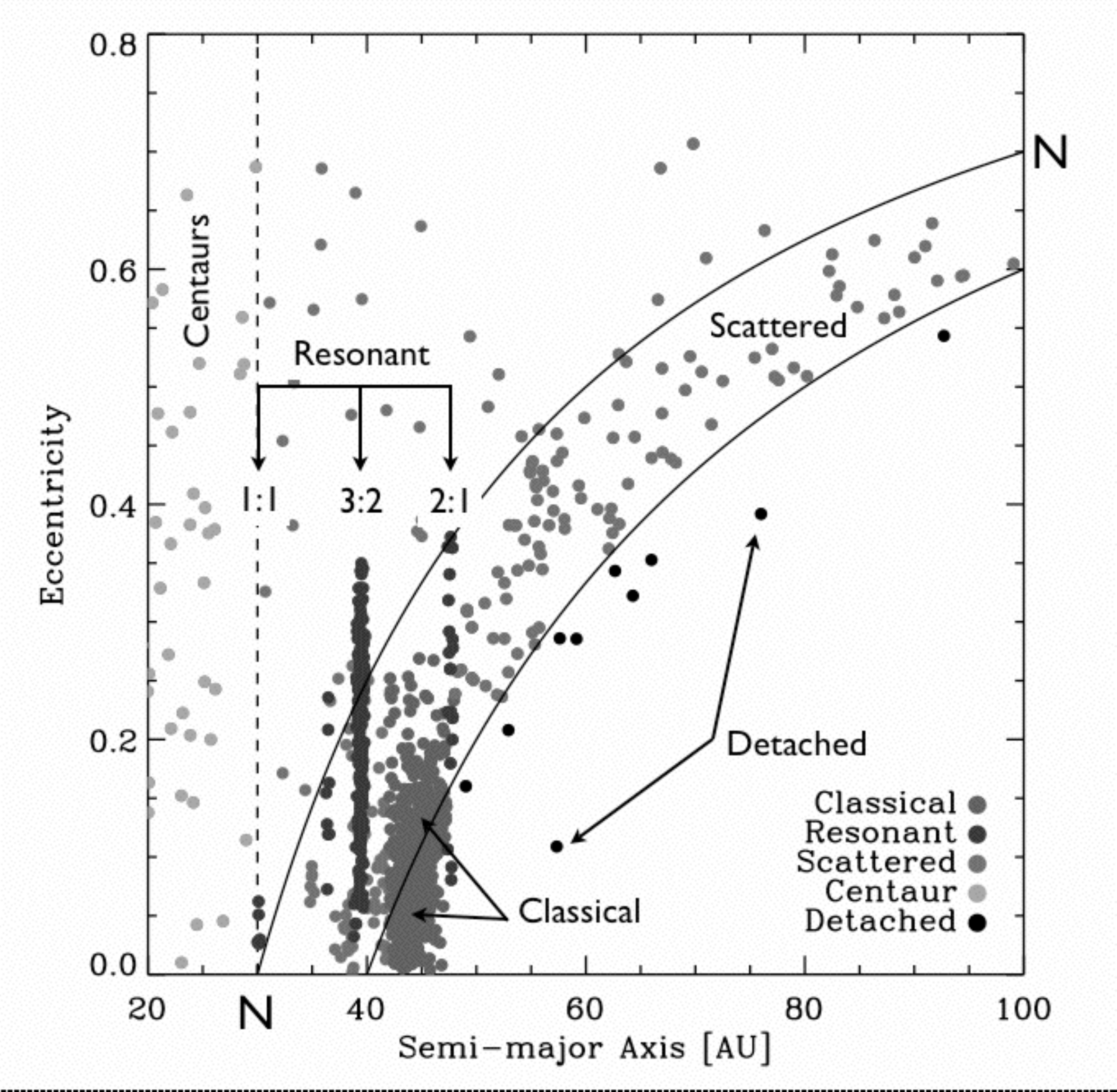} 
\end{center} 
\caption{Semimajor axis vs. orbital eccentricity for the outer Solar system.  The important dynamical sub-types of Kuiper belt objects are
marked on the plot.  Only three of the many populated resonances are indicated.  The locus of orbits having perihelia 
$q$ = 30 AU, equal to Neptune's semimajor axis, is shown as the curve $NN$.  The lower but unlabeled curve shows
the locus of orbits with $q$ = 40 AU.  }
\label{ae_kb}
\end{figure}

The orbits of Kuiper belt objects (KBOs) are empirically grouped into different sub-types (see Figure \ref{ae_kb}).  Most classical KBOs orbit
between about 42 AU and 48 AU and have small eccentricities ($e \lesssim$ 0.2).  They are dynamically stable on Gyr timescales basically because their perihelia never approach the dominant local perturber, Neptune.  Resonant KBOs can cross Neptune's orbit, but are phase-protected from close encounters by their resonant locations and hence can survive on Gyr timescales.  The most famous resonant KBO is 134340 Pluto.  Resonant KBOs provide strong evidence for past planetary migration (Fernandez and Ip 1984, Malhotra 1995). The Scattered KBOs have higher eccentricities ($e \sim$ 0.5, see Figure \ref{ae_kb}) and perihelia in the 30 $\le q \le$ 40 AU range.  Numerical integrations show that the orbits are unstable on Gyr timescales owing to perihelic interactions with Neptune.  These interactions are responsible for kicking up the eccentricities, sending the Scattered KBOs out to large distances (the current record aphelion is 
held by 2000 OO67, at $Q$ = 1300 AU).    The long-term instability of this population means that it is a remnant that has been dynamically depleted over the age of the Solar system, perhaps by factors of a few to ten (Volk and Malhotra 2008).  The Detached KBOs are like the Scattered objects except that their perihelia lie beyond 40~AU, where integrations suggest that Neptune perturbations are unimportant over the age of the Solar system.  The orbits of these objects suggest that another force, perhaps perturbations from an unseen planet (Gladman and Chan 2006, Lykawka and Mukai 2008), from a passing star (Ida et al. 2000), or resonant effects (Gomes et al. 2005) must have lifted the perihelia out of the planetary domain.  The Centaurs, while strictly not part of the Kuiper belt, are thought to be recent escapees from it.  Various definitions exist: the simplest is that they have perihelia \textit{and} semimajor axes between Jupiter (5 AU) and Neptune (30 AU).  The Centaurs interact strongly with the giant planets and have correspondingly short lifetimes, with a median of order 10 Myr (Tiscareno and Malhotra 2003, Horner et al. 2004).

 \subsection{Current Issues}
 
 \begin{itemize}
 
 \item What was the initial mass and radial extent of the Kuiper belt?  
 
 \item How was that mass lost?
 
\item How were the Detached KBOs emplaced?  
 
\item From where in the Kuiper belt do the Jupiter family comets originate?
 
\item Why does the Classical KBO orbit distribution have a sharp outer edge near 48 AU?

\item Did the young Kuiper belt more closely resemble the debris disks of other stars?

\end{itemize}
 
 \section{Main-Belt Comets}
 
  \begin{figure}[t]
\begin{center}
 \includegraphics[width=3.8in]{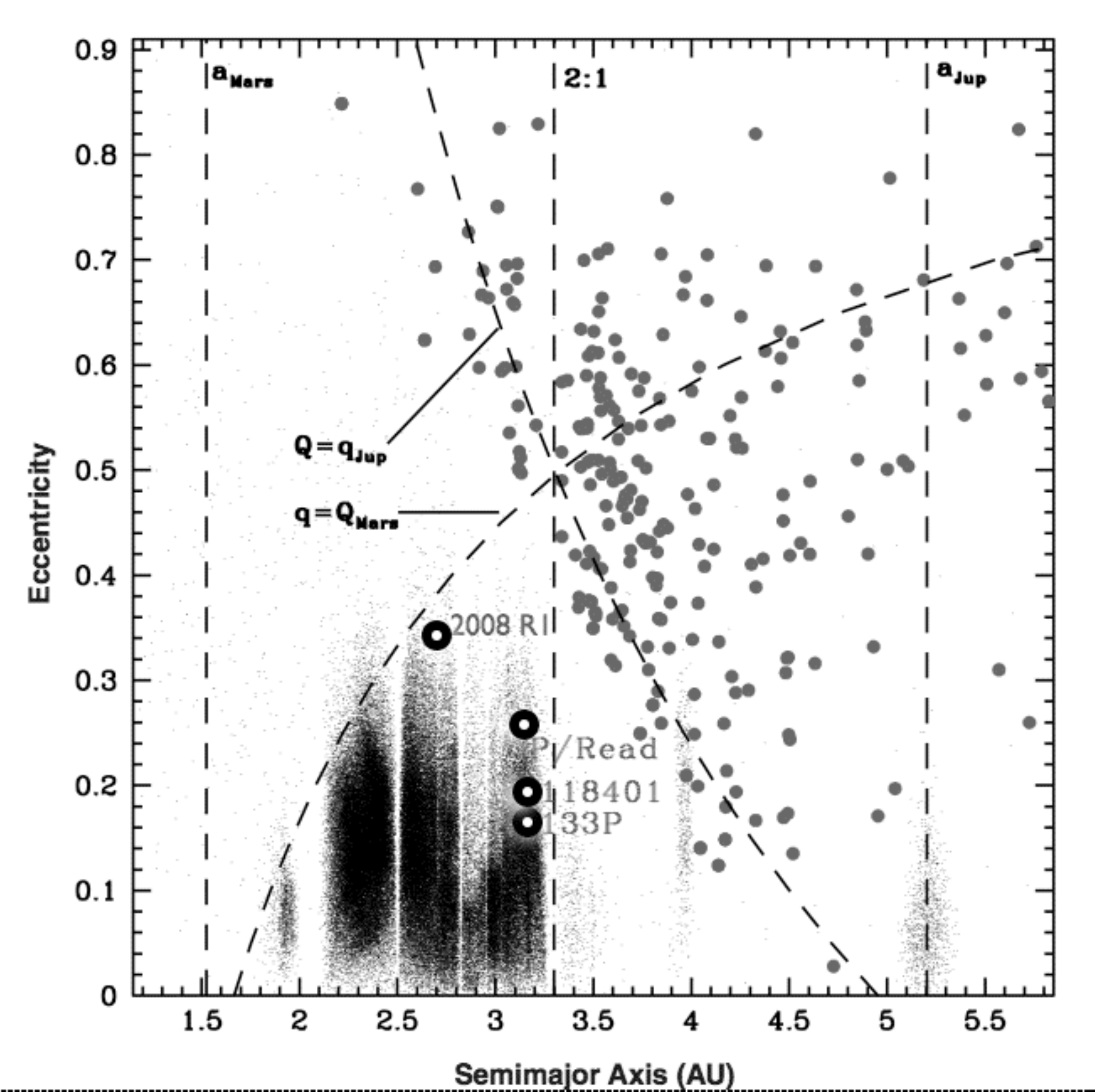} 
\end{center} 
\caption{Semimajor axis vs. orbital eccentricity for the inner Solar system.  The classical comets (mostly short-period comets derived  from the Kuiper belt) are shown as grey circles in the upper right of the diagram.  Main-belt asteroids are represented by single-pixel black points, mostly in the lower left.  The four known MBCs are indicated with large, hollow circles.  Clearly, they occupy the same region of the diagram as is occupied by the main-belt asteroids.  Vertical dashed lines show the semimajor axes of Mars and Jupiter and the location of the 2:1 mean-motion resonance with Jupiter.  Dashed arcs show the locus of orbits that are Mars-crossing (labeled $q = Q_{Mars}$) and Jupiter-crossing ($Q = q_{Jup}$).   }
\label{ae_mbc}
\end{figure}

The Oort Cloud and Kuiper Belt comet reservoirs are relatively well known but the comets of the main-belt represent a still new and relatively unfamiliar population.   The main-belt comets (MBCs) are distinguished from other comets by having asteroid-like ($T_J >$ 3, $a < a_j$) orbits (Hsieh and Jewitt 2006).    In fact, except for their distinctive physical appearances (including comae and tails) these objects would remain indistinguishable from countless main-belt asteroids in similar orbits, as is evident from Figure~\ref{ae_mbc}.

The spatially resolved comae and tails of the MBCs are caused by sunlight scattered from $\sim$10 $\mu$m-sized dust particles.  Several processes might eject grains from the MBCs, but the only one that seems plausible is gas drag from sublimated ice.  For example, a rapidly rotating asteroid might lose its regolith centripetally.  Electrostatic charging of the surface by Solar photons and/or the Solar wind might cause potential differences that could launch particles, as has been observed on the Moon.  But neither of these mechanisms can explain why mass loss from the best-observed MBC (133P/Elst-Pizarro) is episodic and recurrent, being concentrated in the quarter of the orbit after perihelion.  As in many short-period comets at the same distances, gas has not yet been detected; limits to the gas production based on spectroscopic data are of order 1 kg s$^{-1}$ (Jewitt et al. 2009), some 2 to 3 orders of magnitude smaller than in typical near-Earth comets.  
  
\begin{figure}[t]
\begin{center}
 \includegraphics[width=3.8in]{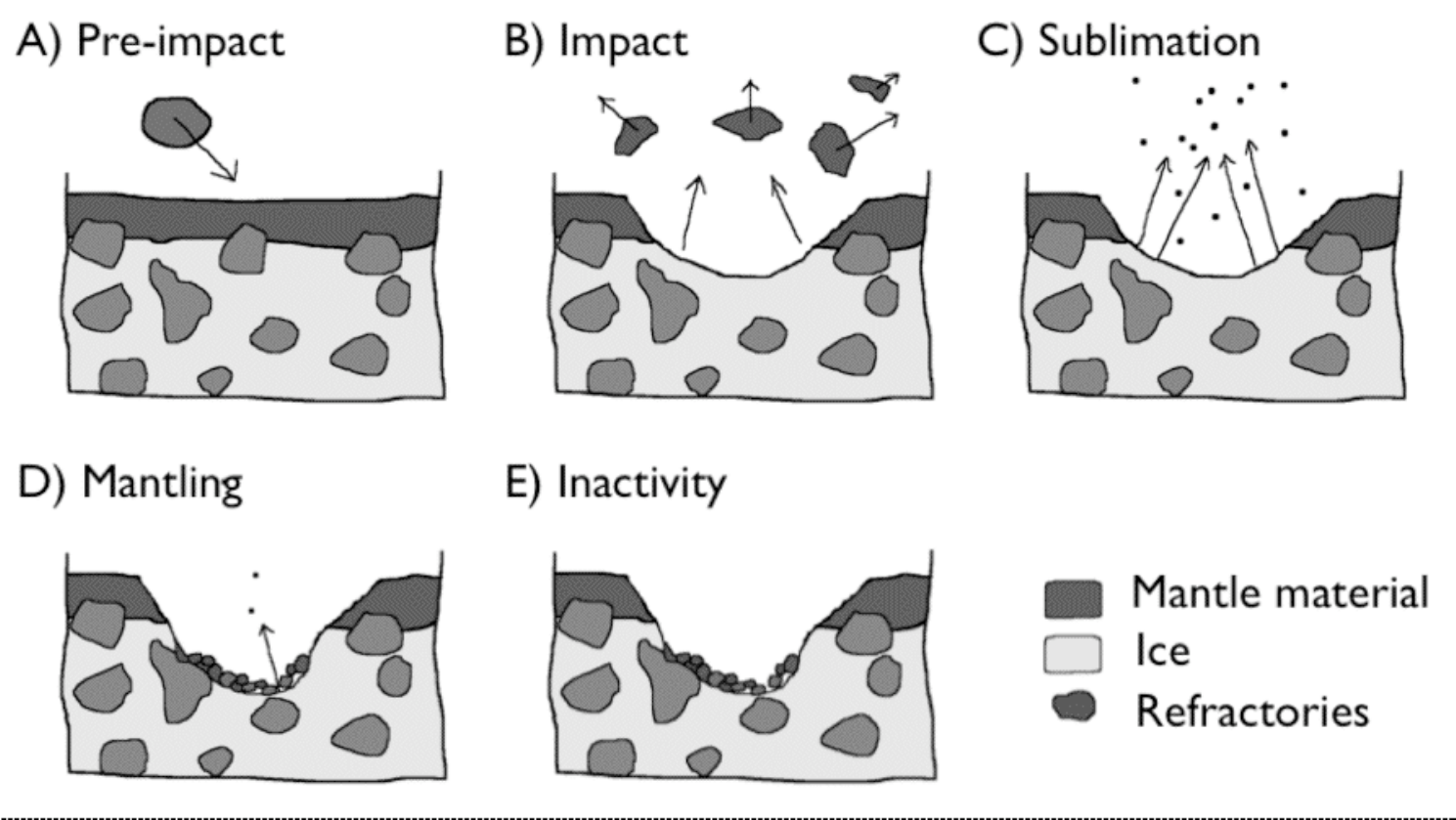} 
\end{center} 
\caption{Conceptual model of the trigger process for main-belt comets. A) The pre-impact surface is mantled in refractory matter (dark layer) beneath which ice (light) and refractory blocks (grey) coexist.  B) Impact of a meter-sized interplanetary boulder blows off the mantle, exposing ice.  C) The ice sublimates, propelling entrained dust into the coma and tail of the comet.  D) Refractory blocks too large to be ejected settle in the bottom of the crater eventually, E), stifling further mass loss.  The process is repetitive with the interval between impacts of meter-sized boulders being perhaps 5,000 to 10,000 yrs.  The stifling of sublimation should be much more rapid, perhaps taking only $\sim$100 yrs of exposure at 3 AU.}
\label{resurface}
\end{figure}

 It is surprising that ice has survived for billions of years in this high temperature ($T \sim$~150 K at $a$ = 3AU) location.
 Stifling of sublimation by a surface refractory mantle of meter-scale thickness is part of the answer (Schorghofer 2008, c.f. Prialnik and Rosenberg 2009).  Sublimation (at $\sim$1 m yr$^{-1}$) should lead to the production of mantles on cosmically very short timescales (100 yr?).  Activation of mass loss then requires a trigger; we suspect that impacts by boulders are responsible (Figure \ref{resurface}), but have no proof.  The interval between sufficient impacts is likely $\sim$10$^4$ yr, giving a duty cycle of $\sim$1\%.
 
\begin{figure}[t]
\begin{center}
 \includegraphics[width=4.2in]{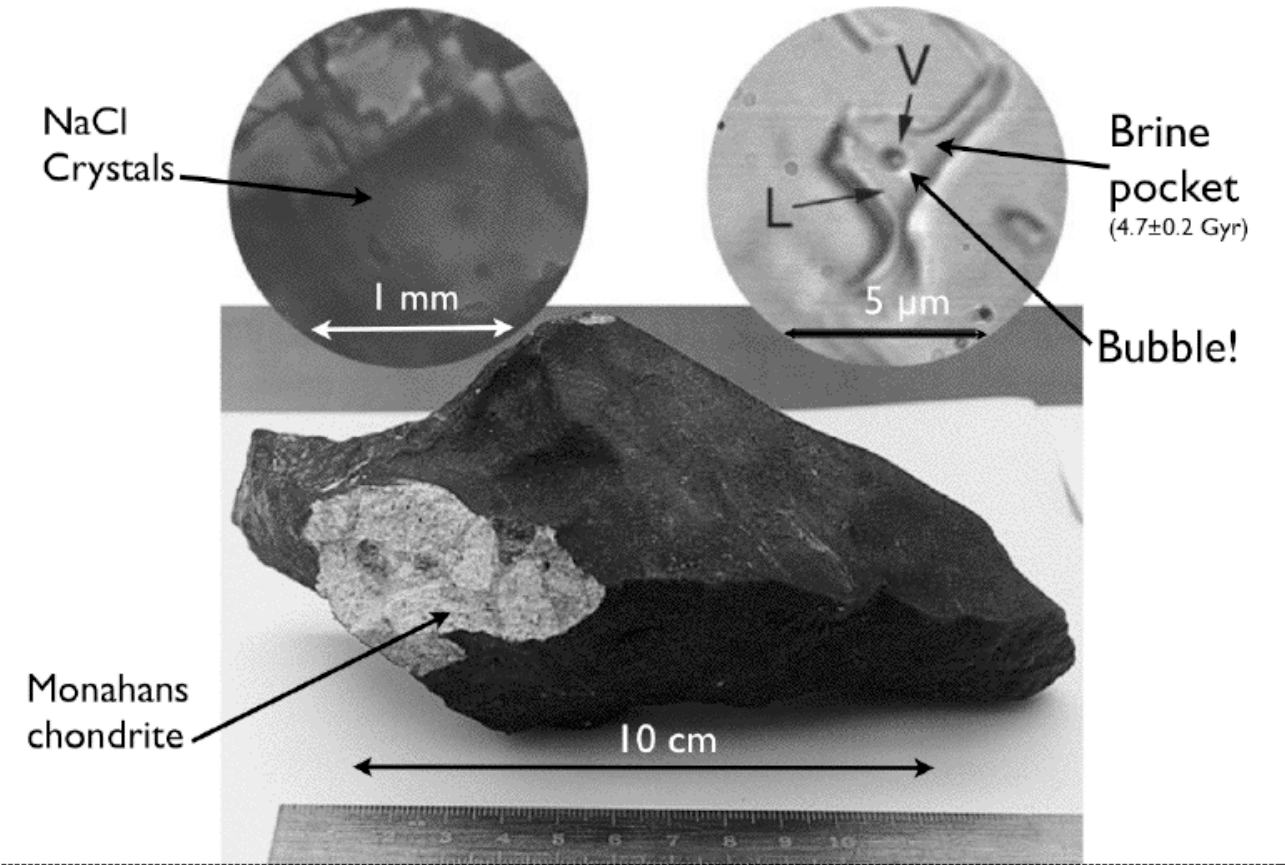} 
\end{center} 
\caption{The incredible H5 chondrite Monahans, in which are found the clear products of aqueous alteration, including cubic salt (NaCl) 
crystals (upper left).  Inside the salt are several liquid brine pockets (upper right) and, inside some of the brine pockets, are gas bubbles.  The halite is radioactively dated to 4.7$\pm$0.2 Gyr.  Monahans is most likely a piece of an outer asteroid belt object in which much of the ice melted soon after formation, and boiled away. It is perhaps closely related to the MBCs.  Figure adapted from Zolensky et al. (1999).}
\label{monahans}
\end{figure}

Evidence for the past presence of water in outer-belt asteroids is strong. Solid-state 
 spectroscopic features that are indicative of hydrated minerals have been found in 10\% or more of 
 asteroids beyond 3 AU. More directly, meteorites from the outer-belt contain hydrated minerals such as serpentine, carbonates
 and halite.  To the author's eyes, the most incredible example is shown in Figure \ref{monahans}, in which is pictured a chondrite meteorite that was observed to fall near the town of Monahans in 
 Texas (Zolensky et al. 1999).  The meteorite fell in a dry place and was collected and immediately handed to scientists for
 study, precluding the possibility of Terrestrial contamination.   Inside the meteorite are found, among other clear
 diagnostics for the past presence of \textit{liquid} water, millimeter-scale cubic halite (NaCl) crystals.  The age of the halite determined from Rb-Sr radioactive decay is  4.7$\pm$0.2 Gyr, essentially equal to the age of the Solar system. Some of these salt crystals further
 contain brine pockets (10 $\mu$m scale) and, in Figure \ref{monahans}, a gas bubble floating in a brine pocket is shown (Zolensky et al. 1999).
   This remarkable specimen leaves no room for doubt about the role of water in the asteroid belt.  The asteroid parent of Monahans contained ice
 that was melted, probably in the first few Myr after the formation of the Solar system.  The liquid water reacted chemically with adjacent minerals forming hydrated species.  Evaporation lead to increasing salinity and the formation of brines in which salt crystals grew.   Then, as $^{26}$Al and other short-lived elements decayed away, the asteroid cooled and any remaining water froze.  A subsequent collision blasted off fragments, one of which struck Texas and appears in Figure \ref{monahans}.

 Watery asteroids are especially interesting in the context of
 the origin of the Earth's oceans and other Terrestrial planet volatiles.   It is likely that the
 Earth accreted at temperatures too high for water to be held by adsorption (indeed, the
 Earth took 50 to 100 Myr to attain final mass and was molten for most of this time).  Instead, volatiles may have been retained only after
 the Earth had cooled to form a surface crust. Impacts of comets and water-containing
 asteroids are the most likely sources of these volatiles.  Recent discussion has moved against delivery
 of water by comets, based
 on the difference between the mean deuterium to hydrogen ratio measured in comets
 ($D/H \sim$ 3$\times$10$^{-4}$) and the ratio in standard mean ocean
 water (SMOW $D/H$ = 1.6$\times$10$^{-4}$) (Meier and Own 1999).  The cometary measurements are, however, of low significance and
 limited to a few comets, with recent evidence suggesting that a wider range might be present.  
 It has been asserted that the asteroid belt represents a more efficient source of water, in the sense
 that impact probabilities from sources in the asteroid belt are higher than from corresponding sources
 in the Kuiper belt (Morbidelli et al. 2000).  Most likely, water is a mix from different sources, but the outer
 asteroid belt is almost certain to be one of these.

Where did the MBCs form?  Occam's Razor suggests that they formed in-place, a possibility that is consistent with
current models and data in which the snow-line in the protoplanetary disk spent some time closer to the Sun than
the current asteroid belt (i.e. asteroids could have trapped ice at formation).  No convincing dynamical pathway from the
Kuiper belt has been identified, given the current layout of the Solar system.  However, if the layout were substantially different
in the past (e.g. because of planetary migration) then it is possible that some objects from the outer regions could
have been trapped in the asteroid belt.

 \subsection{Current Issues}
 
 \begin{itemize}
 
 \item  How many MBCs are there?  Do \textit{all} outer-belt asteroids contain ice?
 
 \item Is the triggered sublimation model of MBC activity (c.f. Figure \ref{resurface}) correct?
 
 \item  How is the ice in MBCs isotopically or compositionally distinct from the ice in comets from the Oort or Kuiper reservoirs?

 \item Did they form locally (as seems most likely) or were they captured from elsewhere?
 
 \end{itemize}
 
 
  \section{The Unknown}
The clear and very uplifting lesson of all this is that we remain only dimly aware even of our immediate astronomical backyard.  The Kuiper belt with its unexpected dynamical subdivisions and structures, the main-belt comets, the (perhaps) densely populated inner Oort cloud, were all completely unknown when, for example, most of the readers of this article were born.  The obvious next question is ``What else is out there?''.  

\begin{table}[b]
  \begin{center}
  \caption{What Else is Out There?$^1$}
  \label{detect}
 {\scriptsize
  \begin{tabular}{|l|c|c|c|c|}\hline 
{\bf Object} & {\bf V(1,1,0)$^2$} & {\bf R$_{24}$[AU]$^3$} & {\bf R$_{grav}$[AU]$^4$}  \\ 
 \hline
Pluto & -1.0 & 320 & --  \\ \\ \hline
Earth & -3.9 & 620 & 50  \\ \\ \hline
Neptune & -6.9 & 1230 & 130  \\ \\ \hline
Jupiter & -9.3 & 2140 & 340  \\ \\ \hline

\end{tabular}
  }
 \end{center}
 \scriptsize{
 {\it Notes:}\\
 $^1$ Modified from Jewitt (2003).\\
  $^2$ The absolute magnitude, equal to the apparent magnitude when corrected to unit heliocentric and geocentric distances and to zero phase angle. \\
  $^3$ The distance at which the apparent red magnitude falls to $m_R$ = 24, the approximate limit of the planned
  Large Synoptic Survey Telescope.\\
  $^4$ The distance beyond which the gravitational influence of the object on the major planets would go undetected.  Pluto has no entry because its gravitational influence is undetectable at any relevant distance.  Calculations are from Hogg et al. 1991 (see also Zakamska and Tremaine 2005).
  }
  \end{table}

To gain some idea of the vast swaths of the Solar system in which substantial objects could exist (even in large numbers) and yet remain undetected, see Table \ref{detect}.  The Table shows that Earth, for example, would appear at red magnitude 24 if displaced to $\sim$600 AU and would certainly be undetected in any existing all-sky survey.  Even ice giants (Neptune in Table \ref{detect}) and gas giants (Jupiter) would be undetected given the present state of our all-sky surveys.  It is widely assumed that unseen massive planets would be detected by their gravitational perturbations of the known planets (or comets).  Table \ref{detect} shows that this is not the case, and optical detection (if taken to magnitude 24 over the whole sky) is the more sensitive method.  While pointing to our extreme state of ignorance concerning the contents of the Sun's potential well, the Table also can be read as a sign of good news.   There is lots of work to do and all-sky surveys to 24th magnitude (and beyond) are both technologically feasible and planned.   All we need now is money.
 
\section{Acknowledgements}
I thank Michal Drahus and Aurelie Guilbert for comments, and NASA's Planetary Astronomy program for support.


\begin{thebibliography}{}

\bibitem[Dones et al.(2004)]{2004come.book..153D} Dones, L., Weissman, 
P.~R., Levison, H.~F., \& Duncan, M.~J.\ 2004, Comets II, 153 

\bibitem[Fernandez 
\& Ip(1984)]{1984Icar...58..109F} Fernandez, J.~A., \& Ip, W.-H.\ 1984, Icarus, 58, 109 


\bibitem[Francis(2005)]{2005ApJ...635.1348F} Francis, P.~J.\ 2005, Ap. J., 
635, 1348

\bibitem[Gladman 
\& Chan(2006)]{2006ApJ...643L.135G} Gladman, B., \& Chan, C.\ 2006, Ap. J. Lett., 643, L135 

\bibitem[Gomes(2009)]{2009CeMDA.104...39G} Gomes, R.~D.~S.\ 2009, Celestial 
Mechanics and Dynamical Astronomy, 104, 39 

\bibitem[Gomes et al.(2005)]{2005CeMDA..91..109G} Gomes, R.~S., Gallardo, 
T., Fern{\'a}ndez, J.~A., 
\& Brunini, A.\ 2005, Celestial Mechanics and Dynamical Astronomy, 91, 109

\bibitem[Gurnett et al.(1997)]{1997GeoRL..24.3125G} Gurnett, D.~A., Ansher, 
J.~A., Kurth, W.~S., \& Granroth, L.~J.\ 1997, Geophys. Res. Lett., 24, 3125 



\bibitem[Higuchi et al.(2007)]{2007AJ....134.1693H} Higuchi, A., Kokubo, 
E., Kinoshita, H., \& Mukai, T.\ 2007, A. J., 134, 1693 

\bibitem[Hogg et al.(1991)]{1991AJ....101.2274H} Hogg, D.~W., Quinlan, 
G.~D., \& Tremaine, S.\ 1991, A. J., 101, 2274 


\bibitem[Horner et al.(2004)]{2004MNRAS.354..798H} Horner, J., Evans, 
N.~W., \& Bailey, M.~E.\ 2004, MNRAS, 354, 798 

\bibitem[Hsieh 
\& Jewitt(2006)]{2006Sci...312..561H} Hsieh, H.~H., \& Jewitt, D.\ 2006, Science, 312, 561 


\bibitem[Ida et al.(2000)]{2000ApJ...528..351I} Ida, S., Larwood, J., 
\& Burkert, A.\ 2000, Ap. J., 528, 351 

\bibitem[Jewitt et al.(1996)]{1996AJ....112.1225J} Jewitt, D., Luu, J., 
\& Chen, J.\ 1996, A. J., 112, 1225 

\bibitem[Jewitt 
\& Luu(2000)]{2000prpl.conf.1201J} Jewitt, D.~C., \& Luu, J.~X.\ 2000, Protostars and Planets IV, 1201 

\bibitem[Jewitt(2003)]{2003EM&P...92..465J} Jewitt, D.\ 2003, Earth Moon and Planets, 92, 465 

\bibitem[Jewitt et al.(2009)]{2009AJ....137.4313J} Jewitt, D., Yang, B., 
\& Haghighipour, N.\ 2009, A. J., 137, 4313 


\bibitem[Kaib 
\& Quinn(2009)]{2009Sci...325.1234K} Kaib, N.~A., \& Quinn, T.\ 2009, Science, 325, 1234 

\bibitem[Kenyon 
\& Luu(1999)]{1999ApJ...526..465K} Kenyon, S.~J., \& Luu, J.~X.\ 1999, Ap. J., 526, 465 


\bibitem[Levison et al.(2002)]{2002Sci...296.2212L} Levison, H.~F., 
Morbidelli, A., Dones, L., Jedicke, R., Wiegert, P.~A., 
\& Bottke, W.~F.\ 2002, Science, 296, 2212 

\bibitem[Lykawka 
\& Mukai(2008)]{2008AJ....135.1161L} Lykawka, P.~S., \& Mukai, T.\ 2008, A. J., 135, 1161 

\bibitem[Malhotra(1995)]{1995AJ....110..420M} Malhotra, R.\ 1995, A. J., 110, 
420 


\bibitem[Marsden and Williams(1999)]{1999cco..book.....M} Marsden, B.~G., \& Williams, G.~V.\ 1999, Catalogue of cometary orbits, 13th ed., by Brian G.~Marsden and Gareth V.~Williams.~Cambridge, Mass.: Central Bureau for Astronomical Telegrams and Minor Planet Center, Smithsonian Astrophysical Observatory, 1999

\bibitem[Morbidelli et 
al.(2000)]{2000M&PS...35.1309M} Morbidelli, A., Chambers, J., Lunine, J.~I., Petit, J.~M., Robert, F., Valsecchi, G.~B., \& Cyr, K.~E.\ 2000, Meteoritics and Planetary Science, 35, 1309 


\bibitem[Moro-Martin et al.(2009)]{2009ApJ...704..733M} 
Moro-Martin, A., Turner, E.~L., and Loeb, A.\ 2009, Ap. J., 704, 733 

\bibitem[Oort(1950)]{1950BAN....11...91O} Oort, J.~H.\ 1950, Bull. Astron. Inst. Neth, 11, 91 

\bibitem[Porco et al.(2005)]{2005Sci...307.1237P} Porco, C.~C., et al.\ 
2005, Science, 307, 1237

\bibitem[Prialnik 
\& Rosenberg(2009)]{2009MNRAS.399L..79P} Prialnik, D., \& Rosenberg, E.~D.\ 2009, MNRAS, 399, L79 

\bibitem[Schorghofer(2008)]{2008ApJ...682..697S} Schorghofer, N.\ 2008, 
Ap. J., 682, 697

\bibitem[Tiscareno 
\& Malhotra(2003)]{2003AJ....126.3122T} Tiscareno, M.~S., \& Malhotra, R.\ 2003, A. J., 126, 3122 

\bibitem[Trujillo et al.(2001)]{2001AJ....122..457T} Trujillo, C.~A., 
Jewitt, D.~C., \& Luu, J.~X.\ 2001, A. J., 122, 457

\bibitem[Vaghi(1973)]{1973A&A....24..107V} Vaghi, S.\ 1973, Astron. Ap., 24, 107 


\bibitem[Volk 
\& Malhotra(2008)]{2008ApJ...687..714V} Volk, K., \& Malhotra, R.\ 2008, Ap. J., 687, 714 

\bibitem[Wiegert 
\& Tremaine(1999)]{1999Icar..137...84W} Wiegert, P., \& Tremaine, S.\ 1999, Icarus, 137, 84 

\bibitem[Wyatt(2008)]{2008ARA&A..46..339W} Wyatt, M.~C.\ 2008, Ann. Rev. Astron. Ap., 46, 339 

\bibitem[Zakamska 
\& Tremaine(2005)]{2005AJ....130.1939Z} Zakamska, N.~L., \& Tremaine, S.\ 2005, A. J., 130, 1939 


\bibitem[Zolensky et al.(1999)]{1999Sci...285.1377Z} Zolensky, M.~E., 
Bodnar, R.~J., Gibson, E.~K., Jr., Nyquist, L.~E., Reese, Y., Shih, C.-Y., 
\& Wiesmann, H.\ 1999, Science, 285, 1377 




\end{thebibliography}
\end{document}